\documentclass[a4paper,preprint,onecolumn,showpacs,amsmath]{revtex4}

\usepackage{graphicx}
\usepackage[dvips]{color}

\begin{document}
\title{Evolution of the gaps through the cuprate phase-diagram}

\author{W. Guyard, M. Le Tacon, M. Cazayous, A. Sacuto}
\affiliation{Laboratoire Mat\'eriaux et Ph\'enom$\grave{e}$nes Quantiques (UMR 7162 CNRS),
Universit\'e Paris Diderot-Paris 7, Bat. Condorcet, 10 rue Alice Domon et Léonie Duquet,75205 PARIS, France}
\affiliation{Laboratoire de Physique du Solide, ESPCI, 10 rue Vauquelin 75231 Paris, France}
\author{A. Georges}
\affiliation{Centre de Physique Th\'eorique, Ecole Polytechnique, 91128 Palaiseau Cedex, France}
\author{D. Colson, A. Forget}
\affiliation{Service de Physique de l'Etat Condens\'{e}, CEA-Saclay, 91191 Gif-sur-Yvette, France}
%\normalsize{$^\ast$To whom correspondence should be addressed; E-mail: alain.sacuto@univ-paris-diderot.fr.}

\date{\today}

\begin{abstract}
The actual physical origin of the gap at the antinodes, and a clear identification of the
superconducting gap are fundamental open issues in the physics of high-$T_c$
superconductors.
Here, we present a systematic electronic Raman scattering study of a mercury-based
single layer cuprate, as a function of both doping level and temperature.
On the deeply overdoped side, we show that the antinodal gap is a true superconducting gap.
In contrast, on the underdoped side, our results reveal the existence of a break point close to
optimal doping below which the antinodal gap is gradually disconnected from superconductivity.
The nature of both the superconducting and normal state is distinctly different on each side of
this breakpoint.
\end{abstract}

%\pacs{74.72.-h, 74.62.Dh, 78.30.-j}

\maketitle
\date{\today}
In conventional superconductors, electrons bind into pairs and minimize their potential energy by a quantity $\Delta$ known as the superconducting gap. Superconductivity arises when pairs are formed and condense in a phase-coherent state below the critical temperature $T_c$~\cite{Tinkham_book}. According to standard BCS theory, the energy $2\Delta$ associated with pair breaking is proportional to $T_c$ and the larger the binding energy, the higher $T_c$ is.

\par
In hole-doped high-$T_c$ cuprate superconductors, $T_{c}$ is strongly dependent on the charge carrier concentration (the doping level). It takes a dome like shape with two distinct regimes (overdoped and underdoped) on each side of the optimal doping
$p ~\sim 0.16$ where $T_{c}$ reaches its maximum. It is now established that the superconducting gap has a $d$-wave symmetry across the
entire phase diagram~\cite{Tsuei_PRL04}, it reaches its maximum values along the antinodal directions and vanishes along nodal directions corresponding respectively to the principal axes and the diagonal in the Brillouin zone.

Among earlier investigations on the superconducting state, angle resolved photoemission spectroscopy
(ARPES)~\cite{Damascelli_RPM07} and electronic
Raman scattering (ERS)~\cite{Venturini_JPCS01,Sugai_PRB03,LeTacon_NaturePhysics06,Devereaux_RMP07} have shown
that the gap at the antinodes, $\Delta_{AN}$, increases continuously as one goes more deeply into the
underdoped side. Although not resolved in momentum, tunnelling spectroscopies also support this experimental
observation~\cite{McElroy_PRL05, Fischer_RMP07}. In contrast, $T_{c}$ changes in the opposite way, in contradiction with the constant ratio $\Delta/T_c$ of standard BCS theory.
More surprisingly, ARPES has revealed that the fingerprint of this antinodal gap persists well above $T_{c}$ in the underdoped regime ~\cite{Damascelli_RPM07,Norman_Nature98}, and disappears only at a much higher temperature scale $T^*$.

%The antinodal gap below $T_{c}$ then merges continuously as a function of temperature
%into another gap like feature above $T_{c}$ which disappears beyond another temperature $T^{*}$.

This persistence of the antinodal gap into the normal state has been named the pseudogap~\cite{Timusk_RPP99}. It is also apparent (although not in a momentum-resolved manner) in Scanning Tunnelling Spectroscopy (STS)~\cite{Fischer_RMP07} and in Nuclear Magnetic Resonant (NMR) experiments~\cite{Timusk_RPP99}, which first revealed this phenomenon.
%
%In contrast, no sign of  pseudo gap is detected by ARPES ~\cite{Norman Nature 98}or NMR ~\cite{Alloul} in the overdoped regime though it is still observed by STS measurements\cite{Renner_PRL98}\cite{McElroy05}\cite{GomesNature07}.

\par
All these experimental observations raise a number of questions.  What is the actual nature and physical origin
of the antinodal gap in the underdoped regime? How to properly identify the  superconducting gap?
Is the normal-state pseudogap a precursor of the superconducting gap or not? The evolution of the antinodal
gap inside and outside the superconducting dome from the overdoped to the underdoped side is therefore one of
the most open and fundamental issues in the physics of high-$T_c$
superconductors~\cite{Norman_adv05,MillisScience07,Hufner_condmat07}.
Our ERS experiments have a direct bearing on these key issues.

We have performed a systematic
study of the antinodal gap as a function of both doping level and temperature, throughout the cuprate phase diagram.
Our experimental findings demonstrate that, in the deeply overdoped regime, the antinodal gap is
clearly associated with superconductivity and can be identified with the superconducting gap.
It has the same doping dependence than the critical temperature $T_c$, its temperature dependence
matches with a standard {\sl d}-wave BCS form, and it disappears at $T_{c}$ leaving a well characterized metallic state.

In contrast, in the underdoped regime, {\it each of these three observations} become invalid.
The antinodal gap becomes very weakly temperature dependent (it even increases slightly as the temperature is
raised), it is no longer proportional to $T_c$, and the ratio $2\Delta_{AN}/k_BT_c$ blows up as the doping level is reduced. Furthermore, the existence of the
antinodal pseudogap well into the normal state and its disappearance at a higher temperature scale $T^*$ is clearly revealed here for the first time by ERS.

%Our detailed mapping of the cuprate phase diagram  allowed us to identify a critical zone in the vicinity of the optimal doping where the antinodal gap becomes progressively very weakly temperature dependent below $T_c$ and the pseudo gap appears above $T_c$.

Our measurements also reveal that, in the underdoped regime, nodal properties become clearly distinct from antinodal ones. In particular, the
characteristic scale associated with the superconducting gap amplitude near the nodes continuously tracks $T_c$.
Our observations strongly support that the antinodal gap is progressively disconnected from superconductivity as one goes from the overdoped to
the underdoped side of the superconducting dome. The nature of the superconducting state is distinctly different on each side of a characteristic doping level which is located close to optimal doping.

ERS measurements have been performed on $HgBa_{2}CuO_{4+\delta }$ ($Hg-1201$) single crystals in a large range of doping levels, from $p = 0.09$ to $p = 0.25$. The $Hg-1201$ single crystals have been grown by the flux method, and oxygen annealing has been carried out in order to overdope the
crystals~\cite{Bertinotti_PhysicaC96}. $Hg-1201$ is a quite ideal cuprate material. It takes a pure tetragonal symmetry without any $Cu-O$ chain contrary to $YBa_{2}Cu_{3}O_{7-\delta}$ (Y-123) or buckling which alters the unit cell of $Bi_2Sr_2CaCu_2O_8$ (Bi-2212). We can then separately measure pure nodal and antinodal responses, without mixing effects. Hg-1201 is made of one single CuO$_2$ layer which is a plane of symmetry in the unit cell. Raman active modes are therefore forbidden in the CuO$_2$ layer. This allows us to investigate the low energy electronic Raman spectrum without being hindered by extra phonons lines \cite{LeTacon_PRB05}. The ERS experiments have been carried out using a triple grating spectrometer (JY-T64000) equipped with a nitrogen cooled CCD detector. The nodal ($B_{2g}$) and antinodal ($B_{1g}$) regions can be explored by using cross polarizations along the Cu-O bond directions and at 45$^o$ from them respectively \cite{Devereaux_RMP07}. In order to track
the antinodal gap we have used the red ($1.9~eV$) laser line which is particularly well adapted for enhancing the Raman signal~\cite{MonsooKang,LeTacon_PRB05}  at low doping levels where the green (2.4 eV) or blue (2.5 eV) excitation lines are unable to detect effectively the signature of the antinodal gap.

\par
In Figure 1, we display the evolution of the anti-nodal Raman responses $\chi ^{\prime \prime}(\omega)$ as
a function of temperature, for four distinct doping levels:
$p = 0.13$, $0.16$, $0.22$ and $p = 0.24$ (Fig. 1 a, b, c, and d respectively) among the fourteen doping levels
that we have studied. There are several experimental facts that we are going to address.
\par
Firstly, for each doping level, (starting from the lowest temperature), we observe a strong antinodal
peak which decreases in intensity as the temperature is raised up to $T_{c}$.
Simultaneously, the low energy continuum grows. This transfer of spectral weight is clearly visible
in the insets of Figure 1. Above $T_c$, a well marked peak is no longer apparent.
\par
Secondly, the antinodal superconducting peak increases in energy as one goes from the overdoped to
the underdoped side. An overview of the evolution of the antinodal peak energy is shown in Fig. 2a
over a large doping range (from $p=~0.25$ to ~$0.12$). The energy of the antinodal peak ($2\Delta_{AN}$)
increases essentially linearly as we enter more deeply in the underdoped regime~\cite{NOTE1}, down to the
lowest doping levels exhibited here.
Such a behaviour has already been observed in previous Raman experiments~\cite{LeTacon_NaturePhysics06,Devereaux_RMP07}.
However our systematic study allows us to clearly point out a drastic change in
the $2\Delta_{AN}(10~K)/k_{B}T_{c}$ ratio as the doping is reduced (see Fig. 2 b).
On the overdoped side, this ratio is nearly constant and takes its smallest values ($\sim 6$),
less than $30\%$ from the value ($\sim 4.3$) expected for a simple {\sl d}-wave BCS gap.
In contrast as one crosses the optimal doping and enters more deeply inside the underdoped regime,
this ratio blows up and doubles $\sim 12$ for $p\sim 0.12$. This reveals a break in the relationship between the antinodal gap and $T_{c}$.
This contrasts with the amplitude of the gap near the nodes, $\Delta_{N}$, measured in the B$_{2g}$ geometry, which continuously follows $T_c$ throughout the superconducting dome (see Fig. 2 a and 2 b).
%The ($B_{2g}$)Raman spectra not shown here will be presented in a next work.
%AG
The ratio $2\Delta_N/k_{B}T_c$ is essentially constant for all doping levels, and equal to
 $\sim 6.4$, to be compared with the d-wave BCS value $\sim 4.3$.
%
%It is then interesting to look more carefully,
Accordingly, the ratio $\Delta_{AN}(T = 10~$K$)/\Delta_{N}(T = 10~$K$)$ has a similar behaviour than the $2\Delta_{AN}(10~$K$)/k_{B}T_{c}$ ratio and exhibits a sharp break. This is shown in Fig. 2 b. This ratio is quasi-constant on the overdoped side and then blows up at optimal doping.
On the overdoped side, its value ($1.1 \pm 0.1$) is in good agreement with the one expected from the antinodal
and nodal Raman responses of a simple {\sl d}-wave BCS superconductor ($\sim 1.2$)~\cite{Devereaux_RMP07}.
\par
Thirdly, Figure 1 reveals that the temperature dependence of the antinodal peak energy changes drastically as the doping level is reduced. For $p=0.24$, it decreases in energy by approximately $50\%$ from $T = 10$~K to $T=40$~K. As the doping level is reduced to $p=0.22$, its energy softening becomes smaller (about 10 percent) and finally for $p=0.13$, the antinodal energy is found to be quasi-constant up to $T_{c}$ (it even appears to slightly increase as the temperature raises). We can then detect an isosbestic point (marked by an arrow in Fig. 1 a) since the superconducting peak energy is quasi constant in this doping range.~\cite{NOTE2}

We therefore note that the temperature dependence of the anti-nodal gap changes in a radical manner as the doping level varies. To quantify this further, we performed a detailed mapping of the antinodal gap as a function of temperature over an extended range of doping levels, thus going further than earlier Raman experiments on $Tl-2201$~\cite{Gasparov_PRB98} or $Bi-2212$~\cite{Staufer_PRL92} compounds. The key point here is to see if there exists an intimate relationship between the clean break in the $\Delta_{AN}/\Delta_{N}(T = 10~$K$)$ ratio measured at low temperature and the antinodal gap evolution at higher temperatures inside the superconducting dome.
In Figure 3, we display the temperature dependence of the anti-nodal peak energy
renormalized to its value at the lowest temperature $T = 10$ K, over a large doping range from $p = 0.24$ to $0.12$.
In the deeply overdoped regime the energy of the antinodal peak is strongly temperature dependent
and matches with the temperature evolution of the BCS d-wave superconducting gap~\cite{Carbotte_PRB95}.
As the doping level is reduced, the temperature dependence of the antinodal gap is weaker
and we can then define a "crossing zone" close to the optimal doping where the anti-nodal peak
energy no longer depends on temperature. In the same doping range, both the $\Delta_{AN}(10~$K$)/k_{B}T_{c}$ and
the $\Delta_{AN}(T = 10~$K$)/\Delta_{N}(T = 10~$K$)$ ratios discussed above blow up.
This points out the strong link which exists between the temperature independence of the antinodal gap
and its splitting-off from the $T_c$ dome.

Such a change in the evolution of the antinodal gap inside the superconducting dome raises
the question of the existence of a break-point in the vicinity of optimal doping. We thus focus now on the normal state in order to see if there exists also a significant change in the antinodal Raman spectra in the normal state on each side of the optimal doping level.
A first glance at Figure 1 immediately reveals that the shape of the "normal" electronic background above $T_{c}$ is
strongly modified as one goes from the overdoped to the underdoped side. The structure of the normal-state electronic
background at low energy changes from a convex to a concave shape.
As we shall show, this drastic change is the Raman signature of the pseudo gap. Up to now, no clear signature
of the pseudo gap in the antinodal Raman response has been yet identified. Most of the earlier Raman
experiments have tracked the pseudo gap in the nodal region ($B_{2g}$ channel)~\cite{Nemetschek_97}.
This seems paradoxical but the main reason is that the $B_{2g}$ Raman signal in the deeply underdoped
regime is much higher than the one in $B_{1g}$ (the antinodal region). There was some attempts to detect
the pseudo gap in $B_{1g}$ channel, but due the presence of weak luminescence contribution,
the spectra were not corrected from the Bose factor at low energy, which prevented these
experiments~\cite{Blumberg_97} to clearly see the evolution of the pseudo gap as a
function of temperature and to identify $T^*$.

Here, we have performed Raman measurements on a slightly under-doped $Hg-1201$ compound
$p = 0.14$ and $T_{c} = 92$~K in order to preserve a significant Raman intensity and avoid spurious
luminescence which occurs at low doping levels. The ERS spectra are shown in Fig. 4 a. The electronic
background just above $T_{c}$ ($T = 95$~K) exhibits a positive curvature.
As the temperature is raised, the low energy electronic background increases in intensity and
we can define a depletion in the measured spectrum
(extending up to $600$ cm$^{-1}$) which "is filling in". This is illustrated in the inset of Fig. 4 a. Above $T=140$~K, the electronic background level remains nearly constant.
%We have then assigned here $T= 140 K$ to the pseudogap temperature $T^{*}$.
Hence, our estimate of the pseudogap temperature for this $p=0.14$ compound is
$T^*\simeq 140$~K.
%AG

This is consistent with the pseudo gap detected by thermoelectric power measurements on $Hg-1201$ \cite{YamamotoPRB_00} as well as with earlier NMR experiments \cite{BobroffPRL_97}.
Our observations are very similar to what is observed by ARPES for a slightly underdoped $Bi-2212$ compound see for instance \cite{Norman_Nature98}. The sharp coherent peak at the antinodes disappears above $T_{c}$ and gives place to a broad feature (the pseudogap) above $T_{c}$. As the temperature is raised, the leading edge midpoint becomes closer to the reference Fermi level and finally it coincides with it between $120$~K and $150$~K. This is in good agreement with our ERS estimate of $T^*$ in underdoped $Hg-1201$ with a $T_{c}=92$~K.
In the same manner, the tunnelling spectra of underdoped Bi-2212 compounds display far above $T_{c}$ a
clear pseudo gap, which manifests itself by a "V" shape feature in the STS spectra and also "fills in" as temperature is raised \cite{Fischer_RMP07}.

\par
On the more overdoped side (see Fig. 1 c or d) there is no sign of pseudogap in the normal state,
and the curvature of the ERS spectrum is always negative.
In Fig 4B, we display the Raman spectra of a slightly
overdoped $Hg-1201$ compound ($p= 0.18$) for several different temperatures above $T_{c}$. The low energy electronic background decreases as the temperature is raised. This corresponds to a conventional metallic behaviour.
It has already been observed in previous Raman experiments~\cite{Venturini_PRL02,Devereaux_RMP07}, where the low-frequency
slope (not fully resolved in Fig.4 b) is proportional to the quasiparticle lifetime and decreases as the temperature is raised.
Our Raman analysis just below and above the optimal doping reveal here unambiguously the existence of the antinodal pseudo gap in the underdoped
side and its disappearance in the overdoped one.

\par
In summary, our experimental findings reveal a clear change of the antinodal Raman response
in both the normal and superconducting states, as one crosses a characteristic doping close to
the optimal doping level. On the strongly overdoped side, the normal state behaves as a rather
conventional metal. The antinodal gap follows $T_c$, it is strongly temperature dependent and closes
at $T_c$ in a very similar way as a BCS d-wave gap. On the underdoped side, the pseudo gap
(the signature of the antinodal gap above $T_c$) is for the first time clearly detected by ERS
and $T^*$ is identified. In contrast to the overdoped side, the antinodal gap in the superconducting state
is no longer (or very weakly) temperature dependent, and it increases continuously as the doping level is reduced, while $T_c$
decreases. This leads to a dramatic increase of the $\Delta_{AN}(10~$K$)/k_{B}T_{c}$ ratio and to a
splitting-off of the antinodal $\Delta_{AN}(10~$K$)$ and nodal $\Delta_{N}(10~$K$)$ gap scales.
Our systematic ERS study of the antinodal gap as a function of both temperature and doping level allowed us to
clearly identify a crossing zone in the vicinity of the optimal doping where all these changes occur.
The physical nature of both the normal and the superconducting state is markedly different on each
side of this characteristic doping.

A clear conclusion of our experimental investigations is that the nature of the antinodal gap changes through
the cuprate phase diagram. It is a true superconducting gap in the deeply overdoped regime, where it gradually
catches up with a d-wave BCS-like description as the doping level increases. In contrast, in the
underdoped regime, the antinodal gap is progressively
disconnected from superconductivity itself (as signaled by the blowing up of the $\Delta_{AN}/k_{B}T_{c}$ ratio).
Our results support the existence of two distinct gaps on the underdoped side of the cuprate phase diagram, as
previously infered from various experiments~\cite{Tallon_PhysicaC2001} such as Andreev-Saint James reflections~\cite{Deutscher_Nature99}, ERS~\cite{LeTacon_NaturePhysics06}, ARPES~\cite{Tanaka,kondo}, or ellipsometry measurements~\cite{Yu_condmat}.

Many different theories have been put forward in order to explain the nature of the antinodal gap in
the underdoped regime (see Refs.~\cite{MillisScience07,Norman_adv05} for a discussion),
and our experimental results put strong constraints on these theories.
One point of view is that on the underdoped side, pairs are preformed in
the normal state below $T^*$ and condense at $T_c<T^*$ (as e.g. on the BEC side of the
BCS to BEC crossover). In this picture, the antinodal gap is associated with the pair
binding (pair breaking) energy. It remains to be seen however whether such models can explain why
a smaller energy scale proportional to $k_BT_c$  shows up in several spectroscopies
when probing the nodal direction (see e.g. Ref.~\cite{Hufner_condmat07} for a recent discussion).
In other theories, the normal-state pseudogap is associated with a phenomenon which is
distinct from pairing. Various possibilities exist: magnetic correlations developing
in the normal state (such as e.g. singlet formation close to a Mott insulator~\cite{anderson}),
or even long-range ordering competing with superconductivity such as orbital
currents~\cite{Varma_PRB97}
or density-wave ordering~\cite{Benfatto_EPJ2000,Chakravarty_PRB01}, possibly disappearing at a quantum critical
point as the doping level is increased.
In such theories, the presence of distinct energy scales associated with the normal-state pseudogap
(dominating the antinodes) and with superconductivity (detectable in the nodal regions)
is a natural possibility. Indeed, recent microscopic calculations based on quantum
cluster methods appear to support a two-gap picture~\cite{Aichhorn_condmat07,civelli_condmat07}
(see also the phenomenological approach of Ref.~\cite{Valenzuela_PRL07}).
Our experimental results do appear easier to interpret within theories in which the
antinodal gap is not associated with superconductivity.
However, such theories have to
cope with the experimental observation that a sharp antinodal peak is only observed
below $T_c$ ({\it not} $T^*$), most probably in connection with the sharpening up of
quasiparticle excitations.
Future experimental investigations and more refined theoretical calculations
have to be performed and compared to each other in order to advance our understanding of these
fundamental issues in the physics of high-temperature superconductivity.

\section*{ACKNOWLEDGEMENTS}
We are grateful to Ph.~Monod, Y.~Gallais, G.~Kotliar and A.~Millis for very helpful
discussions.

\newpage
\begin{figure}
\begin{center}
\includegraphics[width=16cm]{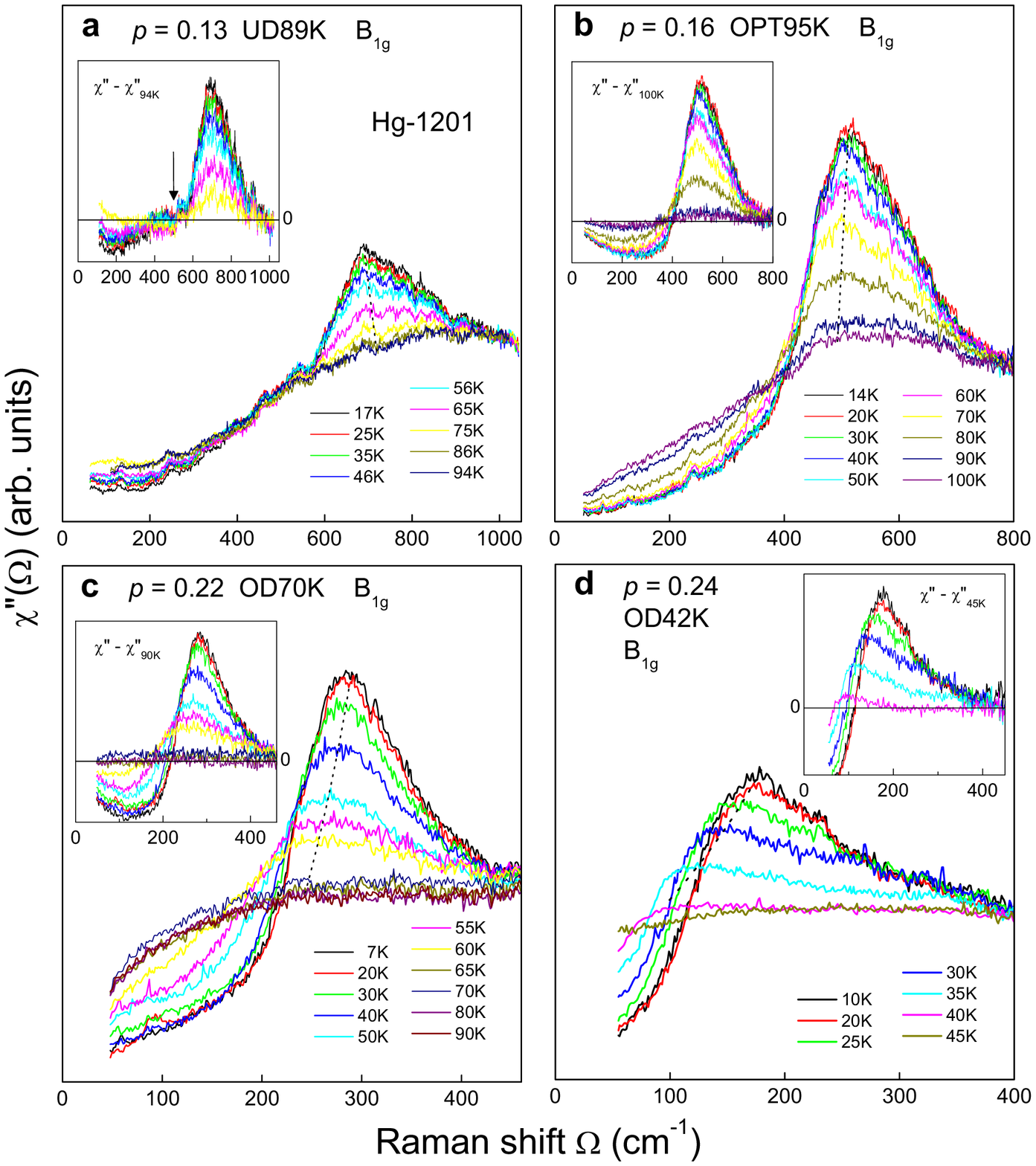}
\end{center}\vspace{-10mm}
\caption{Imaginary parts of the antinodal ($B_{1g}$) Raman response functions $\chi ^{\prime \prime}(\omega)$ as a function of temperature (up to $T_c$) for four doping levels: $p$ = 0.13(\textbf{a}), 0.16 (\textbf{b}), 0.22 (\textbf{c}) and 0.24 (\textbf{d}). The doping value $p$ is inferred from $T_c$ using Tallon's equation \protect\cite{PreslandPhysicaC91}: $1-T_c/95 = 82.6\,(p-0.16)^2$. The dashed lines are guides for the eyes and track the locations of the superconducting peak maxima. The insets exhibit the Raman response substracted from the one just above  $T_c$. The arrow in the inset (\textbf{a}) indicates the isosbestic point.}
\label{fig1}
\end{figure}

\begin{figure}
\begin{center}
\includegraphics[width=10cm]{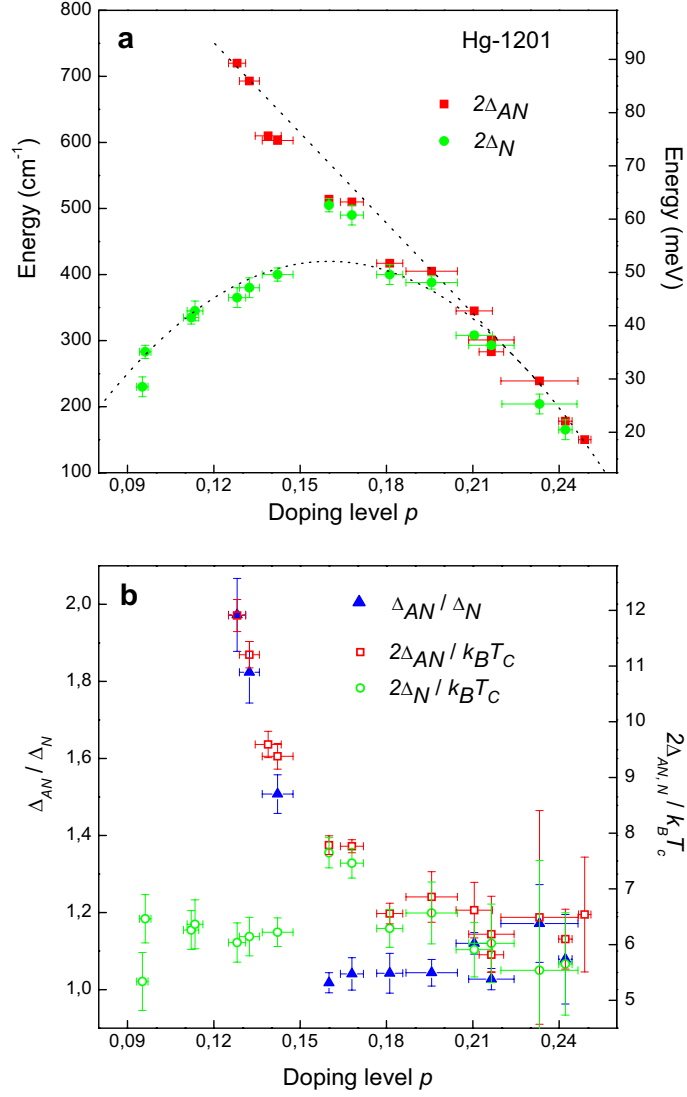}
\end{center}\vspace{-10mm}
\caption{\textbf{a}, Evolutions of the antinodal and nodal gap amplitudes $2\Delta_{AN}$ (red squares) and $2\Delta_{N}$
(green circles) over a large doping range. $2\Delta_{AN}$ and $2\Delta_{N}$ have been defined
from the location of the peaks in the superconducting ERS spectra of $Hg-1201$ with $B_{1g}$ and $B_{2g}$ geometries respectively. The $B_{2g}$ spectra are not shown here. The dome-shaped dashed line displays the energy scale $6.4\,k_BT_c(p)$, with $T_c(p)=0.695*95*(1-82.6*(p-0.16)^2)$ given by Tallon's formula.
The dashed straight line is a guide to the eyes.\textbf{b}, Doping dependence of the $\Delta_{AN}(T = 10$K$)/\Delta_{N}(T = 10$K$)$ ratio deduced from the antinodal and nodal peak locations (blue circles); $2\Delta_{AN}(T = 10$K$)/k_{B}T_c$ ratio (magenta squares); $2\Delta_{N}(T = 10~$K$)/k_{B}T_c$ ratio (green circles).}

\label{fig2}
\end{figure}

\begin{figure}
\begin{center}
\includegraphics[width=14cm]{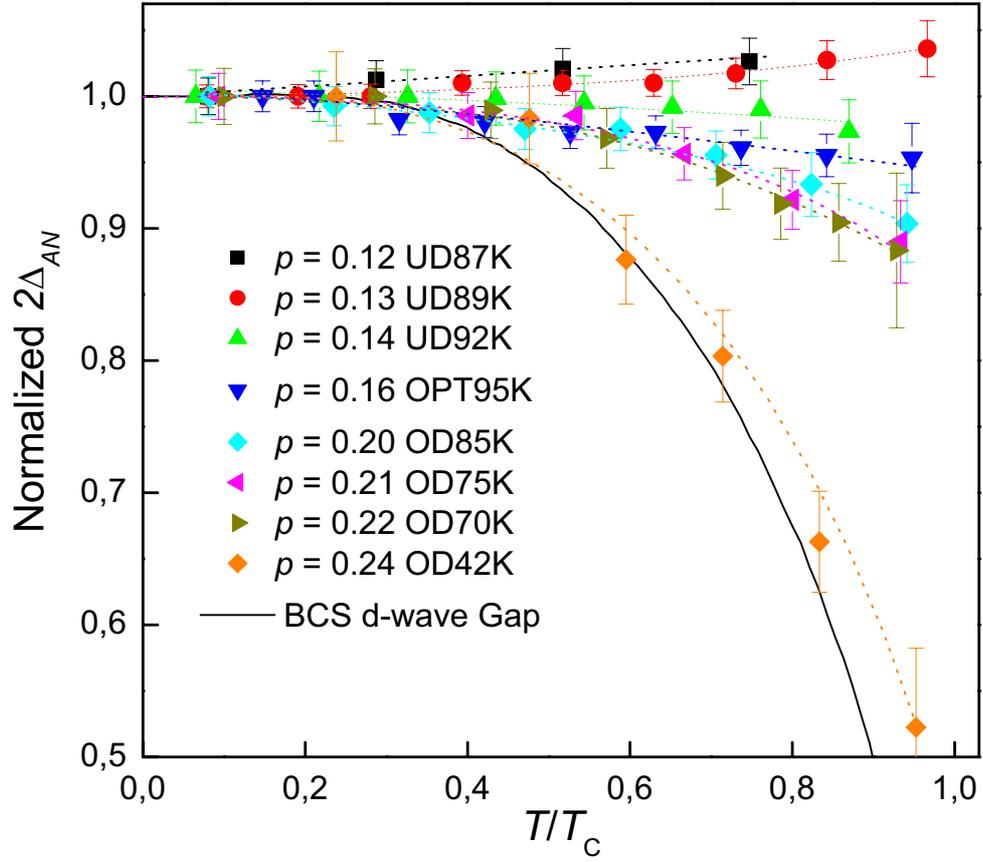}
\end{center}\vspace{-5mm}
\caption{Temperature evolution of the normalized energy of the antinodal superconducting peak $2\Delta_{AN}$ (with respect to its value at $T= 10$~K) for several doping levels. The dashed lines are guides for the eyes. The full black line corresponds to the temperature dependence of a BCS d-wave gap from \protect\cite{Carbotte_PRB95}.}
\label{fig3}
\end{figure}

\begin{figure}
\begin{center}
\includegraphics[width=14cm]{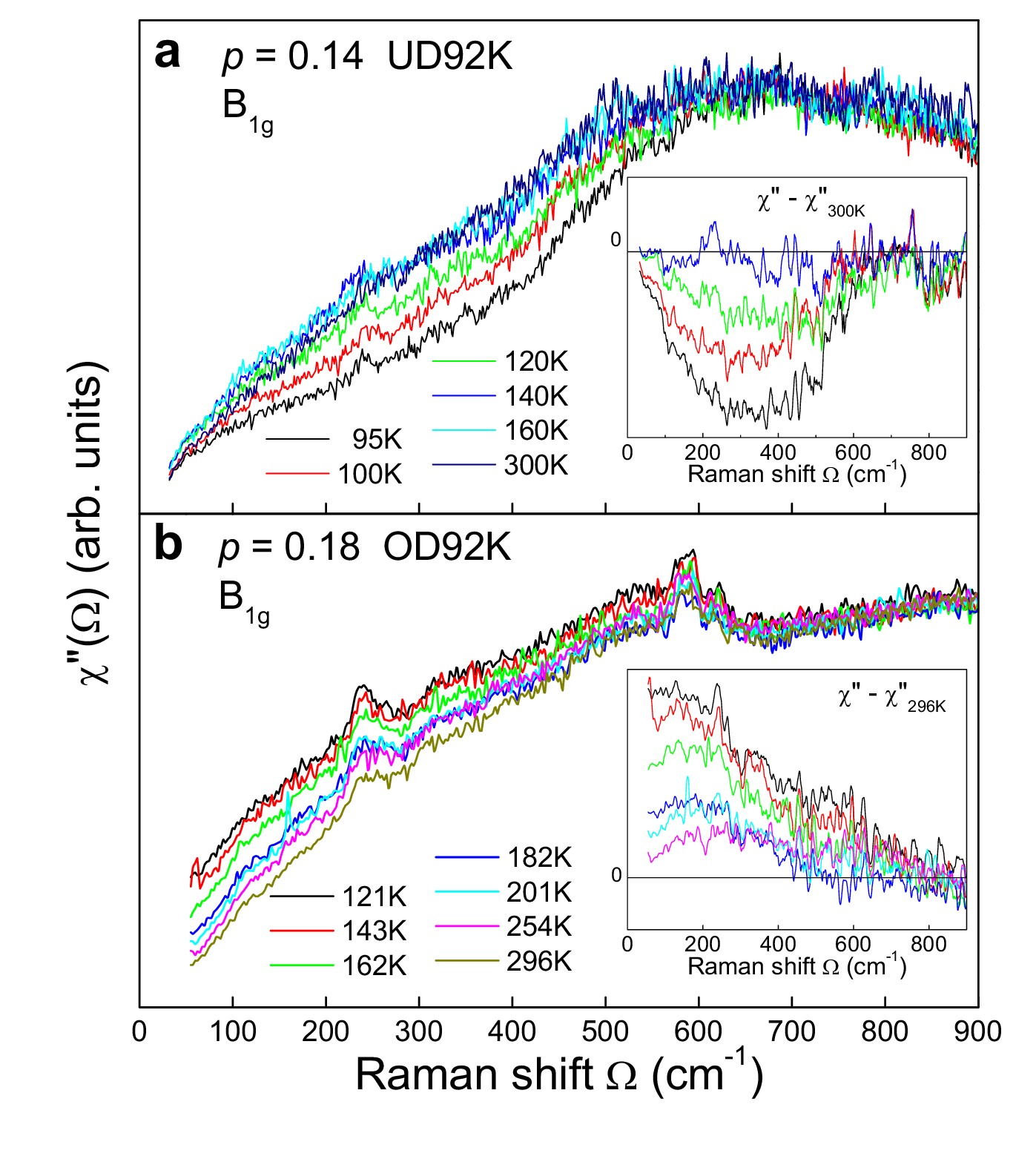}
\end{center}\vspace{-5mm}
\caption{Temperature evolution of the imaginary parts of the antinodal Raman response functions $\chi ^{\prime \prime}(\omega)$ for a sligtly underdoped $Hg-1201$ (\textbf{a}) and a slightly overdoped one (\textbf{b}) above $T_c$. The overdoped sample has been measured using green (2.4 eV) excitation line. The insets exhibit the Raman response substracted from the one at 300 K (\textbf{a}) and 296 K (\textbf{b}).}
\label{fig4}
\end{figure}


\begin{thebibliography}{9}
\bibitem{NOTE1} No more Raman signature of the superconducting gap opening is detected below $p=0.011$~\cite{Gallais_PRB05}.This is consistent with the reduction of the the antinodal quasiparticle spectral weight $Z_{AN}$ and the confinement of coherent quasiparticles near the nodal regions ~\cite{Norman_Nature98}which leads to the disappearance of the superconductivity signature below $T_{c}$ in the antinodal regions.
\bibitem{NOTE2}It is then tempting to look for a sum rule in order to describe the electronic background redistribution. Unfortunately, attempts are until now unsuccessful. If we draw one's inspiration from the Shastry-Shraiman relation~\cite{Shastry_PRL90}, we find that the integrals of the Raman response function divided by the frequency $\int\chi''(\omega)/\omega d\omega$ taken from each part of the isosbestic point are distinct. Even if the Raman response functions of Hg-1201 underdoped compounds are here truncated at 70 cm$^{-1}$, the low energy integral at low temperature are already larger than the higher one.
\end{thebibliography}
\end{document}